# Transport properties of ultrathin $BaFe_{1.84}Co_{0.16}As_2$ superconducting nanowires

Pusheng Yuan[1,2,4], Zhongtang Xu[1], Chen Li[1]  Baogang Quan[3],

Junjie Li[3] , Changzhi Gu[3] and Yanwei Ma[1,4,a)]

[1]*Key laboratory of applied superconductivity, Institute of Electrical Engineering, Chinese Academy of Sciences, Beijing 100190, China*

[2]*Shanghai Institute of Microsystem and Information Technology, Chinese Academy of Sciences, 865 Changning Road, Shanghai 200050, China*

[3]*Beijing National Laboratory for Condensed Matter Physics, Institute of Physics, Chinese Academy of Sciences, Beijing 100190, China.*

[4]*University of Chinese Academy of Sciences, Beijing 100049, China*

**Abstract**

Superconducting nanowire single-photon detectors (SNSPDs) have an absolute advantage over other types of single photon detectors except the low operating temperature. Therefore, many efforts have been devoted to find high-temperature superconducting materials that are suitable for preparing SNSPDs. Copper-based and $MgB_2$ ultra-thin superconducting nanowires have been already reported. However, the transport properties of iron-based ultra-thin superconducting nanowires have not been studied. In this work, a 10 nm thick × 200 nm wide × 30 $\mu$m long high quality superconducting nanowire was fabricated from ultrathin $BaFe_{1.84}Co_{0.16}As_2$ films by a lift-off process. The precursor $BaFe_{1.84}Co_{0.16}As_2$ film with a thickness of 10 nm and root-mean-square roughness of 1 nm was grown on $CaF_2$ substrates by pulsed laser deposition. The nanowire shows a high superconducting critical temperature $T_c^{zero}$=20 K with a narrow transition width of $\Delta$T=2.5 K and exhibits a high critical current density $J_c$ of $1.8\times10^7$ A cm$^{-2}$ at 10 K. These results of ultrathin $BaFe_{1.84}Co_{0.16}As_2$ nanowire will attract interest in electronic applications, including SNSPDs.

---

a)   Author to whom correspondence should be addressed. E-mail: ywma@mail.iee.ac.cn

# Introduction

Superconducting nanowire single-photon detectors (SNSPDs) demonstrate a distinct advantage over other types of single-photon detectors, which have generated tremendous interest in the scientific community [1, 2]. Although SNSPDs have more advantages than other single-photon detectors such as avalanche photodiodes, one drawback is the low operating temperature because current SNSPDs are fabricated from conventional low-temperature superconductors (NbN, WSi). To increase the operating temperature, many efforts have been devoted to explore suitable superconducting materials with high $T_c$ for the preparation of SNSPDs includes $MgB_2$ and cuprate superconductors [3-8]. Iron-based superconductors, as one important member of high temperature superconducting materials, have attracted many efforts to explore its application in the field of large scale current transport [9, 10] and in micro-electronics or nano-electronics applications [11, 12]. However, the transport properties of ultrathin iron-based superconducting nanowires have not been studied because the critical temperature of iron-based superconducting film strongly depends on the thickness. Typically, thinner films present worse superconductivity [13-15] (except for single layer film [16]). For this reason, exploring iron-based superconducting films applications in the field of micro-electronics or nano-electronics only was limited to a relatively thick scale [17-19]. To our knowledge, there is no report on the transmission characteristics of Fe-based superconducting thin films with thickness less than 30 nm, because the 30 nm thick film shows a $J_c$ one order of magnitude lower than the films which thickness more than 70 nm [13]. To explore a possible application of iron-based superconductors to SNSPDs, it is necessary to prepare a high-quality ultra-thin (about 10 nm) iron-based superconducting films. Among the iron-based superconductors, $BaFe_{2-x}Co_xAs_2$ epitaxial films have been extensively investigated because of the higher critical transition temperature, better stability and the easy growth by pulsed laser deposition (PLD) [20-22]. Therefore, $BaFe_{1.84}Co_{0.16}As_2$ (Ba122:Co) was selected to prepare high quality ultra-thin film and to explore the feasibility of Ba122:Co application in SNSPDs.

In this work, we report the first fabrication of superconducting nanowire from PLD prepared Ba122:Co films in 10 nm thick by micro-nanofabrication technologies. From the transport measurement, the obtained 200 nm wide $BaFe_{1.84}Co_{0.16}As_2$ nanowires show $T_c^{zero}$=20

K, transition width $\Delta T$=2.5 K, and exhibit a high critical current density $J_c$ of $1.8\times10^7$ A cm$^{-2}$ at 10 K, suggesting their potential in making iron-based SNSPDs with high operating temperatures.

**Experimental details**

Although MgO, $(LaAl)_{0.7}(SrAl_{0.5}Ta_{0.5})_{0.3}O_3$ (LSAT), $LaAlO_3$ (LAO), $SrTiO_3$ (STO) and $CaF_2$ single-crystalline substrates are often used to prepare Ba122:Co superconducting films, high-performance superconducting films with thickness less than 30 nm have not yet been prepared [13, 21, 23]. These substrates were chosen and tried to prepare high quality ultra-thin Ba122:Co films by PLD. The details of the growth conditions were reported in our previous work [24]. The substrates were cleaned in an ultrasonic bath by using alcohol and acetone for 5 min, respectively, and then the clean substrate was glued onto a stainless steel sample holder by silver paint. The laser (KrF 248 nm) energy was chosen to be 310-320 mJ per pulse with repetition rate of 9 Hz and the distance between substrate and target was kept fixed at 40 mm. A base pressure of $10^{-7}$ Torr was maintained and increased to $10^{-6}$ Torr during the deposition due to degassing. The Ba122:Co films were prepared at 700 °C, before the film deposition the target surface was cleaned by about 1000 laser pulses. The deposition pulse number was varied in the range of 8,00-12,000 and the corresponding thickness was 10–150 nm. After deposition the film was cooled down to room temperature at a rate of 10 °C min$^{-1}$. As the high quality ultra-thin Ba122:Co film was successfully prepared on $CaF_2$ substrate, in order to get superconducting nanowires, we developed a special nanofabrication procedure for our Ba122:Co film (as is shown in Figure 1). First, a thin film of Cr/Au was deposited on the as-prepared Ba122:Co film by electron-beam evaporation technique (Figure 1 (b)). The thicknesses of chromium and gold are 10 nm and 60 nm, respectively, which are adequate to protect the Ba122:Co film from the moisture during subsequent processes. Then, a layer of PMMA electron-beam resist was applied by spin-coating (Figure 1 (c)). A 200 nm wide line was defined by electron beam lithography (EBL) in the PMMA resist (Figure 1 (d)). For optical alignment in the fabrication of nanowire device, the length of lines was defined as 30 $\mu$m. Thereafter, an electron-beam evaporation of 60 nm chromium film was conducted (Figure 1 (e)) and followed with a lift-off process to obtain chromium line (Figure 1 (f)). Next, the pattern of line was transferred to a multilayer film of Cr/Au on Ba122:Co to form Au/Cr/Ba122:Co nanowire (Figure 1 (g)) by Ar ion-beam etching. The reason to choose Cr as etching mask for the fabrication of superconducting nanowires is that the etching rate of Cr is much slower than that of gold in Ar ion-beam etching process. And the 60 nm thickness of chromium can withstand the etching process of Au/Cr/ Ba122:Co multilayer. With the protection of Cr/Au cover layer, our Ba122:Co nanowire becomes compatible with standard device micro-fabrication processes. Finally, a 10 nm thick × 200 nm wide × 30 $\mu$m long Ba122:Co nanowire with four-electrode configuration were fabricated by photolithography, Cr (5 nm)/Au (60 nm) metallization and lift-off. If the light is incident on the nanowire, the current method of preparing nanowires may lead to a decrease in the performance of the SNSPD, due to that the incident photons are absorbed or reflected by the thick metallic multilayer on the nanowires. The way of irradiating nanowires with back incident is often adopted by SNSPD. Fortunately, the substrate $CaF_2$ of Ba122:Co nanowires has a high optical transmittance. To avoid the absorption and reflection of the thick metallic multilayer when the incident is normal, we can irradiate the nanowires from backside with light passing through the substrate $CaF_2$. In addition, we can also try other processes that may be more suitable for preparing single photon detectors as reported in the references [25].Scanning electron microscope (SEM) and atomic force microscope (AFM) were used to obtain the microstructures and the surface topography of samples, respectively. The transport properties of Ba122:Co films and nanowires were measured in four-probe configuration with a physical property measurement system (PPMS).

**Results and discussion**

The thermal relaxation capacity of SNSPD depends on the heat capacity and heat transfer between film and substrate[26]. If we only consider the impact of heat capacity on the thermal relaxation capacity, thinner film generally exhibits faster thermal relaxation. The specific heat capacity of Ba122:Co [27] is larger than $MgB_2$ [28]; therefore, the same thermal relaxation capacity requires Ba122:Co to be thinner relative to $MgB_2$ film (usually about 10 nm or less than 10 nm in thickness). In order to prepare iron-based superconducting thin film suitable for SNSPDs, Ba122:Co films with various thickness were grown on different single-crystalline substrates by PLD. First, the 150 nm thick Ba122:Co films were deposited on the LSAT and LAO substrates. Figure 2 shows the temperature dependence of resistivity for Ba122:Co films. The critical temperature of the films on the LSAT and LAO substrates were $T_c^{onset}$ = 16 K and $T_c^{onset}$ = 14 K, respectively. These values are much lower than the critical transition temperature of the Ba122 : Co target ($T_c^{onset}$ = 26 K). Moreover, previous reports point out that thinner film tend to induce lower critical transition temperature [13]. Therefore, LSAT and LAO substrates are not suitable for the preparation of ultra-thin Ba122:Co superconducting films. Soon after, Ba122:Co films were grown on MgO and STO substrates. Figure 3 shows resistance vs. temperature (R-T) curves normalized by resistance at 300 K (R(300 K)) for thin films with different thickness from 10 to 187 nm. It can be seen that, with the same film thickness, the critical transition temperature of Ba122 :Co thin films grown on MgO and STO substrates is superior to those grown on LSAT and LAO substrates. At the same time, there is a striking evidence that the film thickness strongly affects the critical transition temperature. As shown in Figure 3 (a) and (b), the critical transition temperature of 150 nm thick Ba122 :Co films grown on MgO substrates were $T_c^{onset}$ = 20 K. Unfortunately, complete superconductivity transition were not observed in the film with a thickness below 37.5 nm even if the test temperature is reduced to 5 K. As it can be seen from Figure 3 (c) and (d), the critical transition temperature of Ba122: Co film grown on the STO substrate was higher than the film on the MgO substrate. However, complete superconducting transition was not observed down to 5 K for film with a thickness of 10 nm. The above results indicate that Ba122: Co films grown on MgO substrate and STO substrates are unable to meet the requirements of the SNSPDs. Fortunately, iron-based superconducting films grown on $CaF_2$ substrates tend to exhibit excellent superconducting properties[23, 29-33]. Thus, the $CaF_2$ substrate was chosen to grow ultra-thin Ba122: Co films. It is exciting that the Ba122:Co films on the $CaF_2$ substrates not only have high superconducting transition temperature and narrow transition width, but also show superconductivity in films with thickness much less than those fabricated on other substrates. As shown in Figure 4 (a) and (b), as the thickness of Ba122:Co film was reduced from 150 nm to 10 nm, the critical transition temperature $T_c^{onset}$ decreased from 25.5 K to 23.5 K and $T_c^{zero}$ decreased from 23 K to 20.8 K. For the same thickness, the critical temperature of Ba122:Co film depend strongly on the substrates. The highest $T_c^{zero}$ and the lowest $T_c^{zero}$ for the Ba122:Co films on $CaF_2$ and LSAT substrates, respectively. The effect of the substrate on the critical transition temperature of the Ba122:Co film are consistent with the previous report [34].Therefore, the 10 nm thick Ba122:Co thin film grown on $CaF_2$ substrate is suitable for exploring its application in the SNSPDs field in terms of film thickness and superconducting transition temperature required by SNSPDs.

The superconducting films used to prepare high-performance SNSPD require not only a sufficiently thin thickness but also extremely low surface roughness. In order to investigate the surface morphology and surface roughness of Ba122:Co thin films, AFM studies were performed . Figure 5 (a) shows a 12.5 nm thick Ba122:Co film with a step on $CaF_2$ substrate. The AFM image was clearly divided into the left and right part, which are the surface topography of $CaF_2$ substrate and Ba122:Co film, respectively. The surface topography of the film shows more or less round islands shape, not as smooth as the $CaF_2$ substrate. As shown in Figure 5 (b), the 10 nm thick Ba122:Co film also exhibits the same surface topography as that of the 12.5 nm thick film. The root-mean-square roughness of Ba122:Co film in a 12 μm×12 μm region is 1.05 nm. Figure 5 (c) and (d) exhibit the SEM images of a 200 nm wide × 30 μm long nanowire prepared by a 10 nm thick Ba122:Co film, which indicate that electrodes are in full contact with the nanowire and the nanowire is uniform in width.

In order to check the influence of micro-nanogrid process on the superconductivity of Ba122:Co nanowire, the superconducting transition of nanowire had been compared with that of the Ba122:Co film in Figure 6 (a). The Ba122:Co film shows a sharp superconducting transition at 23.6 K ($T_c^{onset}$) with a narrow transition width of 2.2 K ($\Delta T_c$) .Compared with the film, the nanowire shows a slightly broader transitions width $\Delta T_c$ =2.5 K and a decrease in $T_c^{onset}$ of only about 1.1 K. It can be concluded that Ba122:Co nanowire and film display nearly the same superconducting transition temperature. Therefore, the method used in this study to process the Ba122:Co film into nanowire is appropriate. Figure 6 (b) exhibits the R-T of Ba122:Co nanowire for different test currents. The nanowire exhibits a superconducting transition temperature of $T_c^{onset}$=22.5 K with a narrow transition width of 2.5 K at a measured current of 0.01 mA. As the measured current raises to 0.1 mA, the nanowire exhibits the same superconducting transition as the situation of 0.01mA, but the transition width $\Delta T_c$ increases from 2.5 K to 5 K. This result indicates that the test current strongly affects the transition width of the nanowires. In general, the superconducting transition width of the superconducting material has a significant broadening in the magnetic field. Therefore, we suspect that the increase in the test current will result in the self-field enhancement of Ba122:Co nanowire, which further affects the increase in $\Delta T_c$.

To further characterize the transport properties of the Ba122:Co nanowire, the current–voltage (*I-V*) measurements were performed to determine the critical current ($I_c$) and the critical current density ($J_c$) as a function of temperature. Figure 6 (c) shows the *I–V* characteristics of Ba122:Co nanowire at selected temperatures. As it can be seen, the resistive state of *I–V* characteristics emerges at each test temperature and at small bias current densities, this phenomenon is the same as $FeSe_{0.5}Te_{0.5}$ superconducting nanowire (500 and 800 nm) [35] but different from Ba122:Co micrometer-sized bridges (2.9, 3.5 and 4.7 μm) [17]. As Nappi *et al* reported, the resistive state emerging at low currents in the *I-V* test is due to the depinning (creep flow) of a very limited number of magnetic field lines [35]. Here, the critical current $I_c$ has been defined as the current at which the voltage reaches the value V = 50 μV across Ba122:Co nanowire, because the value of voltage will increase dramatically and *I-V* curves show normal resistance state as it lager than 50 μV. At 10 K, the $I_c$ of nanowire was 0.36 mA, which corresponds to a large $J_c = 1.8 \times 10^7$ A cm$^{-2}$. The $J_c$ value of Ba122:Co nanowire was comparable to that of $La_{1.85}Sr_{0.15}CuO_4$ and $MgB_2$ nanowire[4, 36] at 10 K. At the raised test temperature, the $J_c$ of nanowire decreases gradually as shown in Figure 6 (d). At 16 K, the $J_c$ was $1.1 \times 10^7$ A/cm$^2$; even when the temperature arrives at 20 K, the $J_c$ still remains above $5.0 \times 10^6$ A/cm$^2$, demonstrating excellent current-carrying capabilities of the Ba122:Co nanowire. The inset of Figure 6 (d) shows *I–V* loop characteristics of Ba122:Co nanowire, which exhibits a voltage jump at a critical current ($I_c$) and a small hysteresis behavior which is typical for long superconducting nanowire[37]. The difference between the critical current $I_c$ and the hysteresis current $I_h$ is about 20 μA. It is unfortunate that the voltage switch effect in our nanowire from the superconducting to the normal state is weaker than previous reports [17], which may affect the output signal of SNSPD readout due to smaller voltage switch.

The *I-V* of Ba122:Co nanowires shows flux–flow type behaviors which were also observed in the early $YBa_2Cu_3O_{7-\delta}$ nanowires[25]. However, with the improvement of film quality and nanowire processing technology, the *I-V* flux–flow phenomenon has been almost eliminated in the highly uniform YBa$_2$Cu$_3$O$_{7-\delta}$ and La$_{1.85}$Sr$_{0.15}$CuO$_4$ nanowires [4, 8]. Therefore, it is possible to eliminate the flux–flow of the Ba122:Co nanowire at low currents by adopting a more suitable process for Ba122:Co nanowires and improving the quality of Ba122:Co thin films. In addition, the characteristics of Ba122:Co nanowire superconducting transition shows a high $T_c$, narrow $\Delta T_c$ and large $J_c$. In particular, the $T_c$ of Ba122:Co nanowire reaching 20 K is comparable to MgB$_2$ nanowires[36, 38]. These results will certainly attract attention in the application of superconducting electronics and may open the door for developing Ba122:Co SNSPDs with higher operating temperatures, although in this case a new patterning process must be developed and the physical mechanism responsible for the *I-V* curve be investigated.

**Conclusions**

For the first time, the 10 nm thick Ba122:Co films were successfully prepared by PLD. Furthermore, we succeeded in developing Ba122:Co nanowires fabrication processes and fabricating superconducting nanowires in 200 nm wide × 30 μm long on Ba122:Co films. The nanowires show $T_c^{zero}$=20 K, $\Delta T_c$=2.5 K and $J_c$=1.8×10$^7$ A cm$^{-2}$ at 10 K. Based on the current results, we believe that the ultra-thin Ba122:Co nanowires will attract interest in electronic applications, including SNSPDs.

**Acknowledgements**

The authors would like to express their thanks to Dr. He Huang，Chao Yao and Chiheng Dong for many useful discussions. This work is partially supported by the National Natural Science Foundation of China (Grant Nos. 51320105015 and 51607174), the Beijing Municipal Science and Technology Commission (Grant No. Z141100004214002), the Beijing Training Project for the Leading Talents in S & T (Grant No. Z151100000315001).

**Captions**

Figure. 1 Schematic diagrams to illustrate the process of nanowire fabrication on ultrathin Ba122:Co films.

Figure. 2 (a) shows the resistance versus temperature of Ba-122:Co films on LSAT and LAO substrates. (b) The detail of superconducting transition for Ba-122:Co films on LSAT and LAO substrates. The resistance has been normalized to the value of R(300 K).

Figure.3 (a) and (c) show the resistance versus temperature of Ba-122:Co films on MgO and STO substrates, respectively. (b) and (d) The detail of superconducting transition of Ba-122:Co films on MgO and STO substrates, respectively. The resistance has been normalized to the value of R (300 K).

Figure. 4 (a) shows the resistance versus temperature of Ba-122:Co films on $CaF_2$ substrates. (b) The detail of superconducting transition for Ba-122:Co films on $CaF_2$ substrates. The resistance has been normalized to the value of R(300 K).

Figure. 5 (a) Atomic force microscope images of the $CaF_2$ substrate and 12.5 nm thick Ba-122:Co film. (b) Atomic force microscope images of 10 nm thick Ba-122:Co film with root-mean-square roughness 1.05 nm in a 12 μm×12 μm region.(c) SEM image of 10 nm thick × 200 nm wide × 30 μm long Ba122:Co nanowire. (d) The enlarged view of Ba122:Co nanowire.

Figure. 6 (a) shows the resistance versus temperature of 10 nm thick Ba-122:Co films and 200 nm wide Ba-122:Co nanowire. Inset: the detail of superconducting transition for Ba-122:Co film and nanowire. (b) The detail of superconducting transition for Ba-122:Co nanowire under different test currents.(c) *I–V* characteristics of 5 nm thick ×200 nm wide × 30 μm long Ba-122:Co nanowire at various temperatures. (d) The temperature dependence of critical current density of the Ba-122:Co nanowire. Inset: *I–V* loop characteristics of Ba122:Co nanowire and hysteresis behavior at 10 K.

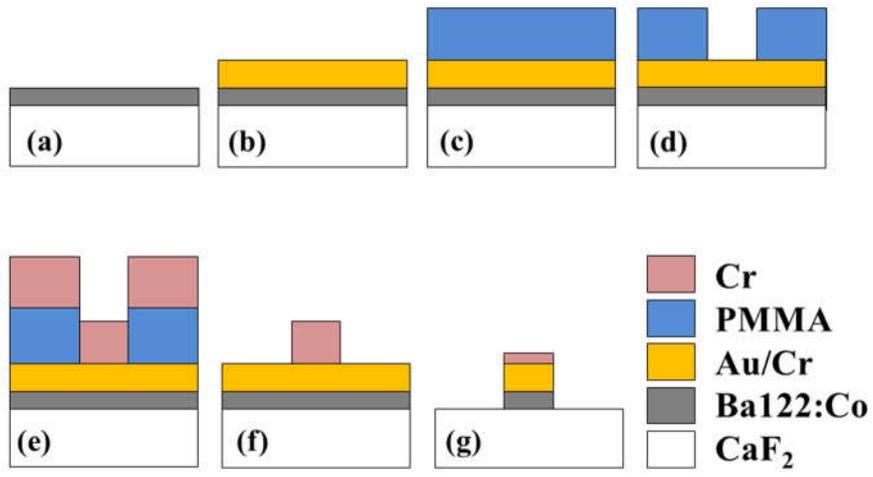

Figure. 1. P.S.Yuan et al.

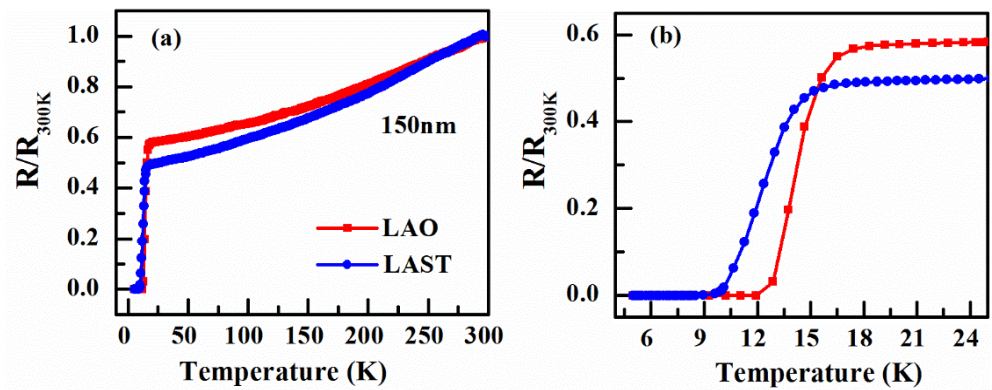

Figure. 2. P.S.Yuan et al.

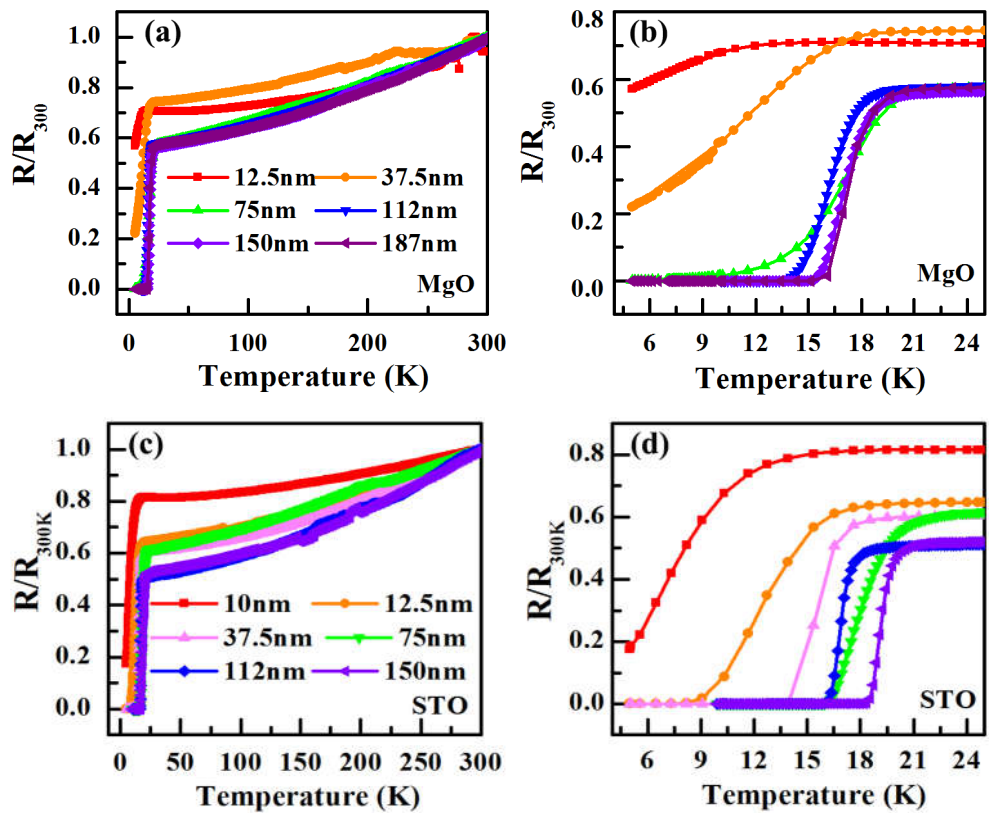

Figure. 3. P.S.Yuan et al.

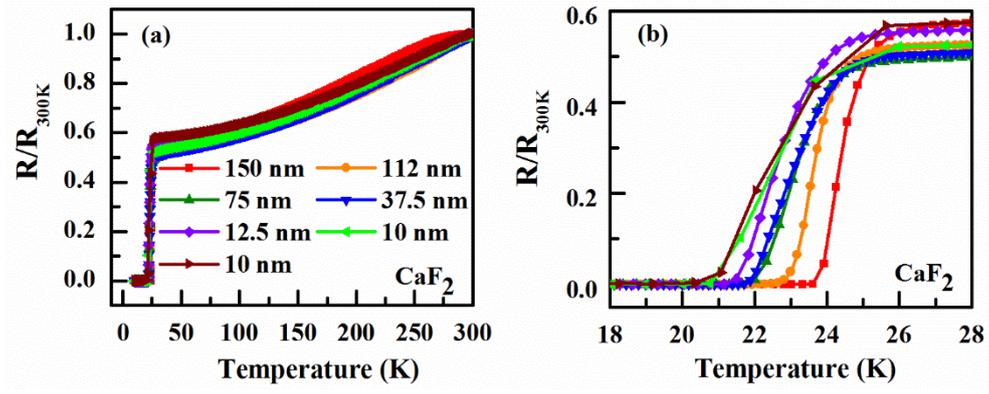

Figure. 4. P.S.Yuan et al.

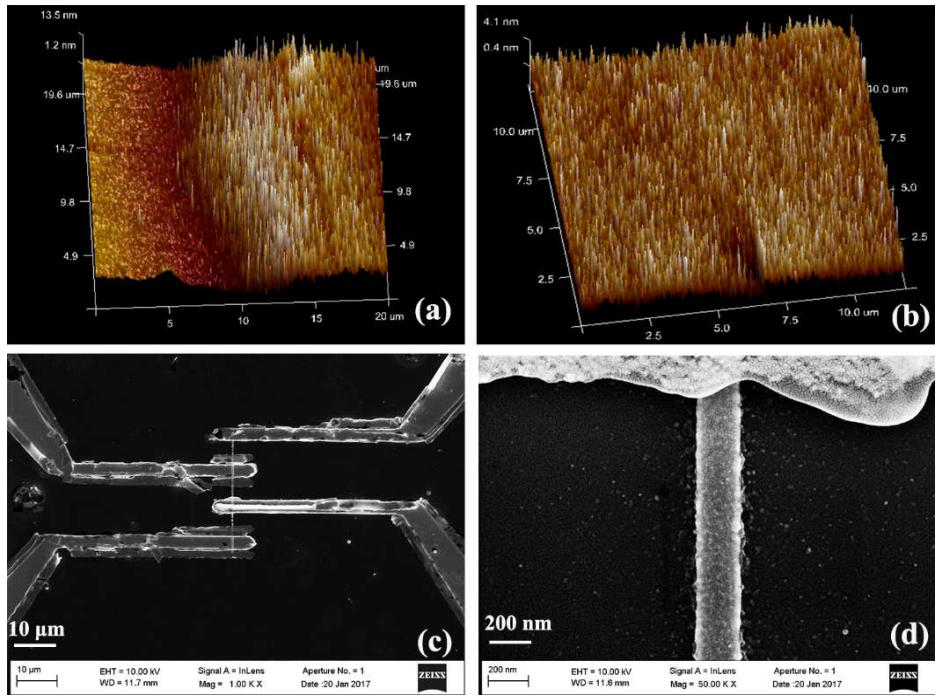

Figure.5. P.S.Yuan.

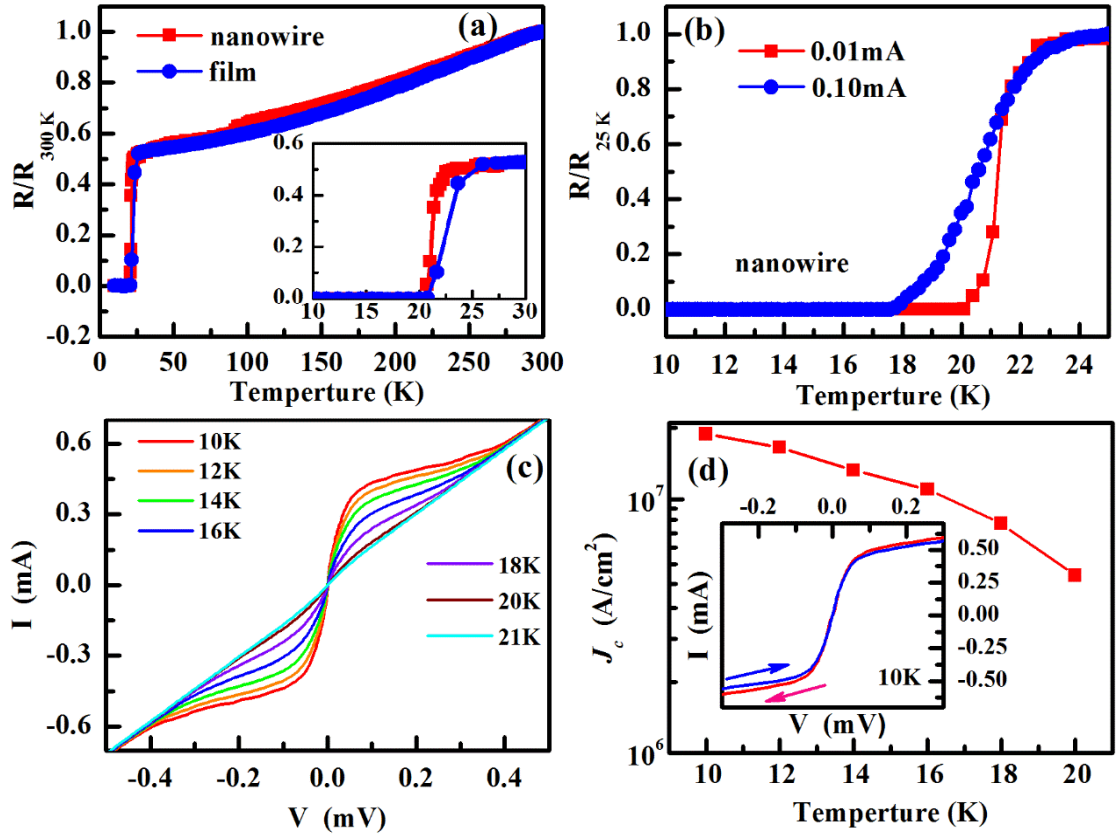

Figure.6. P.S.Yuan.